\documentclass[aip,amsmath,amssymb,reprint]{revtex4-2}
\usepackage{natbib}

\usepackage{amsmath}
\usepackage{amssymb}
\usepackage{bm}
\usepackage{latexsym}
\usepackage{graphicx}
\usepackage{epsfig,epsf,rotating}
\usepackage{indentfirst}
\usepackage{epstopdf}
\usepackage{xcolor}
\usepackage{multirow}
\usepackage{tabularx}
\usepackage{ctable}
\usepackage{balance}

\usepackage{upgreek}
\newcolumntype{C}{>{\centering\arraybackslash}X}
\usepackage{nicefrac}

\begin{document}

\title{Equipartition and the temperature of maximum density of TIP4/2005 water$^\dag$}
\author{Dilipkumar N. Asthagiri}
\affiliation{Oak Ridge National Laboratory, One Bethel Valley Road, Oak Ridge, TN 37830-6012}
\email{asthagiridn@ornl.gov}
\author{Thiago Pinheiro dos Santos}
\affiliation{Oak Ridge National Laboratory, One Bethel Valley Road, Oak Ridge, TN 37830-6012}
\author{Thomas L. Beck} 
\affiliation{Oak Ridge National Laboratory, One Bethel Valley Road, Oak Ridge, TN 37830-6012}

\begin{abstract}
We simulate TIP4P/2005 water in the temperature range of 257~K to 318~K with time-steps $\delta =$ 0.25, 0.50, 2.00, and 4.00~fs. The density-temperature behavior obtained using 0.25 or 0.50~fs are in excellent agreement with each other but differ from those obtained using time-steps that have been shown earlier to lead to a breakdown of equipartition.  The temperature of maximum density (TMD) is 277.15~K with $\delta t = 0.25\;\mathrm{or}\; 0.50$~fs, but is shifted to progressively lower values for longer time-steps, a trend that holds for different thermostat/barostat combinations.  Enhancing the water-water dispersion interaction, as has been recommended for simulating disordered proteins in TIP4P/2005, degrades the description of the liquid-vapor phase envelope. A key takeaway from this study is that using sufficiently short time-steps ($\leq 0.5$~fs) to preserve equipartition is essential for obtaining meaningful liquid water properties and for producing reliable data to parametrize biomolecular simulation models, as correct-ensemble sampling is fundamental to ensure reproducibility across codes and simulation algorithms. 
\newline

\noindent \footnotesize{$^\dag$Notice:  This manuscript has been authored by UT-Battelle, LLC, under contract DE-AC05-00OR22725 with the US Department of Energy (DOE). The US government retains and the publisher, by accepting the article for publication, acknowledges that the US government retains a nonexclusive, paid-up, irrevocable, worldwide license to publish or reproduce the published form of this manuscript, or allow others to do so, for US government purposes. DOE will provide public access to these results of federally sponsored research in accordance with the DOE Public Access Plan (https://www.energy.gov/doe-public-access-plan}

\end{abstract}

\maketitle

\begin{figure}[h!]
\includegraphics[alt={Graphical abstract}]{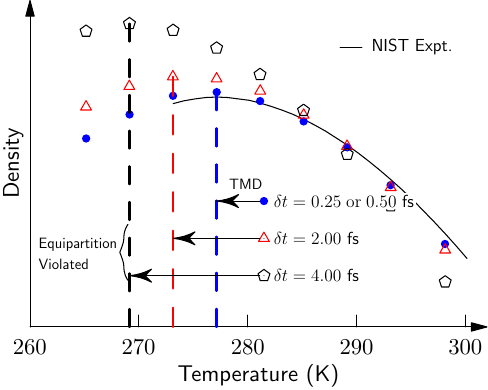}
{Graphical abstract}
\end{figure}

The liquid-to-solid transition of most liquids is accompanied by an increase in density. Water is a notable exception. Liquid water has a density maximum at about 4$^\circ$C (277.15~K)\cite{wagner:iapws99}. For temperatures below 4$^\circ$C, the density of the liquid decreases up to the freezing point, and the solid phase (ice Ih) has a lower density than the liquid. This allows ice to float on water and is critical for life on planet Earth. 

Capturing the temperature of maximum density (TMD) is a sensitive  test of the computer simulation of liquid water\cite{guillot:jml02}.  In this letter we restrict our attention to classical statistical mechanical simulations, which is of first interest in the simulation of large biomolecular systems. It helps to remember, however, that the TMD shows a strong nuclear quantum effect: the TMD of D$_2$O is about $11.6^\circ$C and that of T$_2$O is about $13.4^\circ$C \cite{nist,goldblatt:t2o}. 

It is well known that the otherwise excellent three-site model SPC/E shows a density maximum deep in the supercooled regime ($T < 240$~K) \cite{guillot:jml02}. The four-site TIP4P model improves the TMD to around 250~K \cite{guillot:jml02}.  Abascal and Vega re-parametrized the TIP4P model specifically targeting the TMD, besides including the enthalpy of vaporization, the densities of the liquid at ambient conditions and densities for select ice phases in the parametrization process\cite{vega:tip4p}. They used $NpT$ Monte Carlo simulations on a small system comprising 360 water-molecules to show that the resulting TIP4P/2005 model captures several of the properties of the liquid very well.  They estimated a TMD of 278$\pm$3 (2$\sigma$)~K from a polynomial fit to simulated data.
In a later study from the same group\cite{pi:tip409}, the authors found the TMD of TIP4P/2005 water to be $277\pm 3$~K using molecular dynamics simulations with a time-step of 2~fs, once again using a polynomial fit to identify the TMD. In these and other studies, and to the best of our knowledge,  the TMD is obtained at either 1 bar or 1 atm rather than the saturation pressure. This approximation is acceptable because the Poynting pressure correction is negligible for liquid water for the conditions of interest. We will follow this practice here. 


In an effort to better describe the ensemble of expanded conformations of intrinsically disordered proteins, Shaw and coworkers noted that water models typically used in simulations underestimate the dispersion interaction \cite{piana:water2015}. To this end, they re-parameterized the TIP4P/2005 water model by strengthening the dispersion interactions with reference to quantum chemical calculations while adjusting the partial charges to match target properties. From molecular dynamics simulations at 1~bar and a time-step of $\delta t = 2.0$~fs, they note that the TMD of TIP4P/2005 water is 275~K and that of the re-parameterized TIP4P-D model is 270~K.

 Recently, we have shown that the $\delta t$ used in molecular dynamics simulations of liquid water has non-negligible impacts on the computed properties. In particular, we have shown that $\delta t > 0.5$~fs leads to the breakdown of equipartition in the SPC/E and TIP3P models of water \cite{asthagiri:jctc2024a,asthagiri:cs2025}. Ultimately, this breakdown can be traced to the requirement of using a small time-step for capturing the fast rotational relaxation of water, a point that Rahman and Stillinger \cite{rahman:jcp71,stillinger:jcp74} had already noted in their early simulations of liquid water. We showed that the breakdown of equipartition changes the $p$-$V$ behavior, and in particular, for the liquid phase a time-step that breaks equipartition also leads to a larger volume for a given target pressure. Motivated by these observations, here we map the liquid-vapor coexistence curve for the  TIP4P/2005 and TIP4P-D water models in simulations that  obey equipartition.  We find that for choices of time-steps that break equipartition --- and these are also most often used in extant biomolecular and aqueous phase simulations --- the TMD is progressively shifted to lower values as the time-step is increased. 
 
We simulate a system comprising 32,768 water molecules at an initial mass density 980.264 kg/m$^3$ using the NAMD code\cite{phillips2005scalable,namd:2020}. The temperature was maintained using the canonical stochastic velocity rescaling (CSVR) thermostat\cite{svr:jcp07} and the pressure of 1~bar was maintained using a Monte Carlo barostat\cite{chow:mcbarostat1995,aqvist:mcbarostat2004}. The coupling time for the thermostat was 1~ps and the volume moves were attempted every 40~fs (160~steps for $\delta t = 0.25$~fs, 80~steps for $\delta t = 0.5$~fs, and 20~steps for $\delta t = 2.0$~fs, respectively). For computational efficiency in the GPU-enabled simulations, for $\delta t = 4.0$~fs we persist with volume sampling every 20~steps. As shown earlier\cite{asthagiri:cs2025}, with adequate sampling the volumes are well converged with this and
other choices of volume sampling frequency.  Throughout electrostatic interactions were calculated using particle mesh Ewald summation with a PME grid spacing of 1~{\AA} and the tolerance for electrostatic energy at the real-space cutoff being $10^{-7}$, a factor of 10 tighter than the default in the NAMD code. The cutoff distance for real-space electrostatic and LJ dispersion interactions is 10~{\AA}.  We included analytical corrections for long-range LJ interactions.  To test the sensitivity of thermostat/barostat combinations, using GPU-resident simulations we also modeled the phase envelope using the Langevin thermostat/barostat \cite{feller:jcp95}. 

We also performed a limited set of calculations with the Langevin thermostat/barostat on CPUs to ensure that mixed precision calculations do not introduce further uncertainties.   As further checks, we simulated the system with the AMBER\cite{amber1,amber2}, GROMACS\cite{gromacs2,gromacs3}, and OpenMM\cite{openmm,openmm8} codes with different thermostat/barostat combinations. The simulations span single precision arithmetic (in GROMACS) to mixed precision arithmetic (in AMBER and OpenMM). These results are collected in the SI. For time-steps that ensure equipartition, good agreement between all the codes  is obtained for the equilibrium density-temperature behavior. However, this is no longer the case for conditions that violate equipartition, acutely emphasizing 
that ensuring equipartition is necessary for reproducibility across codes and simulation algorithms. 

The sampling lengths are documented in the SI. Here we note that for all simulations, data was logged every 500 steps, and statistical uncertainties were calculated using the Friedberg-Cameron approach \cite{friedberg:1970,allen:error}. 

\begin{figure}
\includegraphics{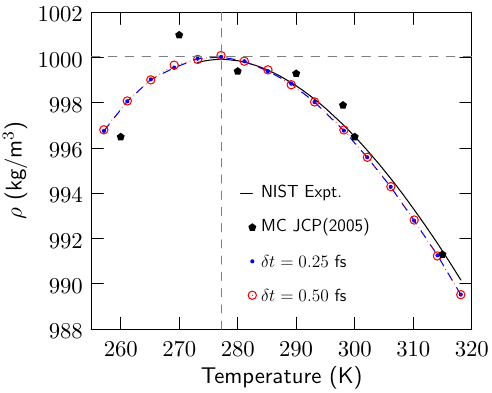}
\caption{Density versus average simulation temperature of TIP4P/2005 water for conditions of most interest in biomolecular simulation. The average simulation temperature equals the setpoint temperature to within 0.02~K.  The dashed vertical line is the experimental TMD at 277.15~K. The horizontal dashed line is a guide to the eye to highlight that within the temperature resolution of this study, the simulated TMD is at 277.15~K. The (black) solid line is the experimental density of the stable liquid under saturation conditions\cite{nist}. The black filled circles are data obtained from Table~1 Ref.\ \citenum{vega:tip4p}; the data is expected to have an uncertainty of $\pm 1$~kg/m$^3$. The size of the symbol for $\delta t = 0.25$~fs is $\approx 2$ standard error of the mean.}
\label{fg:figure1}
\end{figure}

 Figure~\ref{fg:figure1} shows that the TIP4P/2005 model captures the TMD in excellent agreement with experiments. 
 The average of the simulated density with $\delta t = 0.25$~fs and $\delta t = 0.50$~fs is $1000.07\pm 0.03$~kg/m$^3$, which is about 0.01\% higher than the experimental value of 999.98~kg/m$^3$\cite{wagner:iapws99}.  The excellent agreement between the $\rho$-vs-$T$ behavior for the two indicated time-steps for integrating the equations of motion emphasizes that if care is taken to ensure equipartition\cite{asthagiri:jctc2024a,asthagiri:cs2025}, the results are insensitive to the time-step. 

The simulated maximum density here can be compared to the 1000.5~kg/m$^3$ quoted in the original TIP4P/2005 study\cite{vega:tip4p}, a value that was obtained from a 5$^\mathrm{th}$-order polynomial fit to the simulated $\rho$-vs-$T$ data. The inevitable scatter due to the limitation of simulating a small system for a necessarily constrained length and the high estimated uncertainty of $\pm 1$~kg/m$^3$ in the original Monte Carlo simulations of the TIP4P/2005 model undoubtedly cause this discrepancy. 

\begin{figure}[h!]
\includegraphics{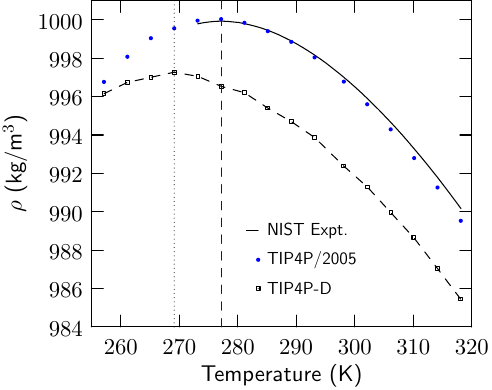}
\caption{Impact of enhancing the water-water dispersion interaction in TIP4P/2005. The time-step for integrating the equations of motion is $\delta t = 0.25$~fs. Rest as in Figure~\ref{fg:figure1}.}
\label{fg:figure3}
\end{figure}
Figure~\ref{fg:figure3} shows that  enhancing the dispersion interaction in TIP4P/2005 as suggested in Ref.\ \citenum{piana:water2015} worsens the predicted $\rho$-vs-$T$ behavior. The TMD of TIP4P-D is shifted to 269.15~K, which can be compared to 270~K reported in Ref.\ \citenum{piana:water2015}. The density at the TMD of $997.27\pm 0.08$~kg/m$^3$ is also lower, being in error relative to experiments by about 0.3\%. These results encourage reconsidering the 
hypothesis that it is the weaker water-water dispersion interaction that is the source of the problems in capturing the conformations of intrinsically disordered peptides within the Amber99SB-ILDN protein force-field.  We showed earlier using theory and computer simulations\cite{tomar:jpcb16,asthagiri:gly15,tomar:jpcl20} that protein-water hydrophilic (attractive) interactions are decisive in the solution thermodynamics and conformational behavior. These hydrophilic interactions counter the drive provided by both primitive hydrophobic hydration of the polypeptide and protein intramolecular interactions that favor the polypeptide chain to collapse. Juxtaposing those findings with the observations in Ref.\ \citenum{piana:water2015} and the results noted above, it appears that
enhancing the water-water dispersion interaction serves to enhance the protein-water hydrophilic interaction to compensate for potentially 
stronger polypeptide intramolecular interactions predicted by the Amber99SB-ILDN protein force-field.

\begin{figure}[h!]
\includegraphics{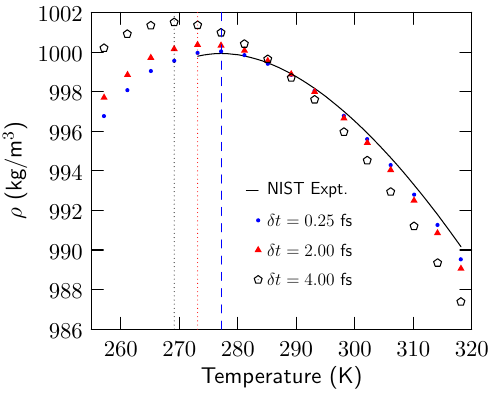}
\caption{Time-step dependence of the predicted saturation curve for two different time-steps. The vertical lines indicate the location of the TMD for $\delta t = 0.25$ (dashed blue), 2.00~fs (dotted red), and 4.0~fs (dotted black). Rest as in Figure~\ref{fg:figure1}.}
\label{fg:figure2}
\end{figure}

Having established the expected behavior in simulations that obey equipartition, we next consider what happens when the time-step is large enough to break equipartition. Figure~\ref{fg:figure2} shows that the breakdown of equipartition shifts the phase envelope:
with $\delta t = 2.0$~fs the TMD is 273.15~K and with $\delta t = 4.0$~fs it is 269.15~K. The shift in magnitude of the TMD because of the breakdown
of equipartition is already comparable to the magnitude one finds experimentally due to nuclear quantum effects. This identification once again emphasizes the sensitivity required in modeling liquid water. The motion of the hydrogen atoms bonded to the parent oxygen are what case the fast librational motions that require a small time-step; it is the same hydrogen atom that is the basis for the strong isotope effect. 

As noted before, Pi et al.\ \cite{pi:tip409} obtained the TMD of TIP4P/2005 to be 277$\pm$3~K from molecular dynamics simulations 
using $\delta t = 2.0$~fs. As Fig.~S3 (SI) shows their $\rho$-vs-$T$ data are in fair agreement with what we compute using 2~fs, data which we know does not obey equipartition. As Fig.~S3 (SI) shows it is important to get both the location of the TMD and the density of the fluid, conditions that are not jointly satisfied for 2~fs. 

\begin{figure*}[ht!]
\includegraphics{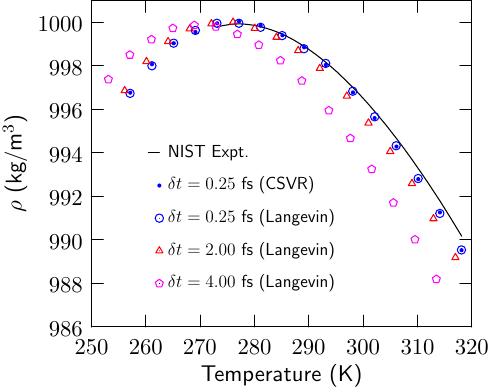}\hspace{5mm}\includegraphics{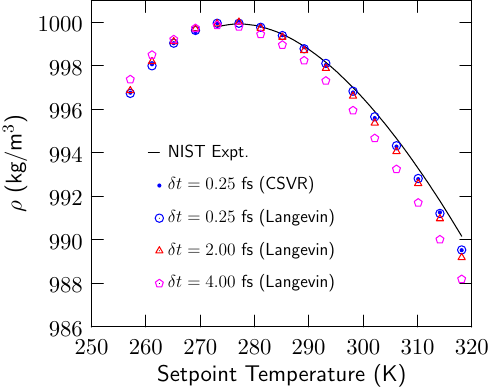}
\caption{Sensitivity of density temperature behavior to thermostat/barostat combination. Data from the stochastic velocity rescaling thermostat/MC barostat  (CSVR) at the smallest time-step is shown together with the results from Langevin thermostat/barostat (Langevin). The $2\sigma$ standard error of the mean is approximately the size of the filled symbol (CSVR). \underline{Left panel}: $\rho$ versus the average simulation temperature. \underline{Right panel}: $\rho$ versus the setpoint temperature defined in the simulation configuration file.}\label{fg:langevin}
\end{figure*}
Figure~\ref{fg:langevin} shows the sensitivity of the results to the choice of thermostat/barostat, specifically CSVR/Monte Carlo barostat versus Langevin thermostat/barostat. We first note that for the smallest time-step used, the results based on the different thermostat/barostat combination are in excellent agreement. (See also figures S1 to S4.) This is as it should be for simulations that obey equipartition and are well converged.  This conclusion is also consistent with 
Fig.~1 of Ref.\ \citenum{asthagiri:cs2025}.  Importantly, the temperature set in the simulation configuration file, the setpoint temperature, is in excellent agreement with the average temperature obtained from the simulation log file.  For longer time-steps, however, the results obtained using the Langevin thermostat/barostat combination reveal additional subtleties. Consider first the left panel (Fig.~\ref{fg:langevin}), where we show the results versus the average temperature obtained from the simulation log file. It is immediately clear that for longer time-steps the average temperature is shifted to lower values: the deviation from the setpoint increases with the time-step. As we noted in Figure~1 of Ref.\ \citenum{asthagiri:jctc2024a}, the Langevin thermostat coupled only to the oxygens (in the rigid TIP4P/2005 model) affects translational and rotational temperatures  to different extents, unlike the CSVR thermostat that affects both modes nearly equally. This then leads to a shift in the average simulation temperature in Langevin simulations with time-steps that break equipartition. But the density versus the setpoint temperature masks the problem in the simulation. This identification is critical in the context of replica exchange simulations where violation of the canonical condition can lead to nontrivial artifacts \cite{hummer:replicathermostat2009}: using setpoint versus average simulation temperatures in replica exchange simulations that do not obey equipartition can lead to confounding results. 

In recent years the idea of hydrogen mass repartitioning\cite{hopkins:hmr2015} has gained much prominence in biomolecular simulations. In this approach the mass of the  hydrogen atoms in biomolecules is increased with the mass of the parent heavy atom appropriately reduced, and the system simulated in liquid water (H$_2$O) with a time-step of 4~fs. Our earlier results\cite{asthagiri:jctc2024a,asthagiri:cs2025} and the present work show that this choice of a time-step is 
ill-advised for proper equilibrium statistical mechanical simulation of water, and hence, for solutes dissolved in water. 

In conclusion, by ensuring equipartition is obeyed in simulations, we find that the TIP4P/2005 model captures the TMD in excellent agreement with experiments. This model also gives a good description of the saturation curve for the stable liquid in conditions of most interest in biomolecular simulations. Our study shows that 
for reproducible prediction of the phase envelope and thus also the thermodynamics of hydration across different thermostat/barostat combinations and different simulations codes and algorithms, it is a foundational requirement to ensure that the simulations obey equipartition. This observation is fundamental in tuning and developing force fields for classical statistical mechanical simulations of aqueous systems. To wit, including a dispersion correction in the TIP4P/2005 model to better predict the behavior of intrinsically disordered peptides, degrades the performance of the liquid itself, suggesting that the flaws in the protein force-field are being compensated by changes in the water model.  Librational motions are a crucial part of the physics of liquid water, and the reason for the fast librations is also why liquid water shows a strong isotope effect: the light hydrogen atoms anchored to a heavy parent oxygen. Thus much sensitivity needs to be exercised in the classical statistical mechanical simulation of water under ambient conditions. 

\section{Acknowledgements}
We thank Rohan Sridhar Adhikari for helpful discussions. This research used resources of the Oak Ridge Leadership Computing Facility at the Oak Ridge National Laboratory, which is supported by the Office of Science of the U.S. Department of Energy under Contract No. DE-AC05-00OR22725. This research used resources of the National Energy Research Scientific Computing Center (NERSC), a Department of Energy User Facility using NERSC award ERCAP0034320.
 
\bibliography{refs.bib}

\end{document}


\title{Supporting Information: Equipartition and the temperature of maximum density of TIP4/2005 water}
\author{Dilipkumar N. Asthagiri}
\email{asthagiridn@ornl.gov}
\affiliation{Oak Ridge National Laboratory, One Bethel Valley Road, Oak Ridge, TN 37830-6012}
\author{Thiago Pinheiro dos Santos}
\affiliation{Oak Ridge National Laboratory, One Bethel Valley Road, Oak Ridge, TN 37830-6012}
\author{Thomas L. Beck}
\affiliation{Oak Ridge National Laboratory, One Bethel Valley Road, Oak Ridge, TN 37830-6012}

\begin{absolutelynopagebreak}

\maketitle

\end{absolutelynopagebreak}

\tableofcontents

\clearpage
 
\section{Methodology}

We simulate a system comprising 32,768 TIP4P/2005 water molecules, initially occupying a cube of volume $100\times 100\times 100$ {\AA}$^3$ (mass density of 980.264 kg/m$^3$). (For simulations with OpenMM, we use a system with 4096 water molecules in a cube of volume $50\times 50\times 50$~{\AA}$^3$.) Electrostatic interactions were calculated using particle mesh Ewald summation. The cutoff distance for real-space electrostatic and LJ dispersion interactions is 10~{\AA}. We include analytical corrections for long-range LJ interactions. 

\subsection{NAMD simulations}

For simulations with NAMD,  the $NpT$ ensemble is simulated using either (1) the canonical stochastic velocity rescaling (CSVR) thermostat\cite{svr:jcp07} + Monte Carlo barostat\cite{chow:mcbarostat1995,aqvist:mcbarostat2004} or (2) the Langevin thermostat+barostat combination \cite{feller:jcp95}. The PME grid spacing is set to  1~{\AA} and the tolerance for electrostatic energy at the cutoff length is $10^{-7}$, a factor of 10 lower than the default in the NAMD code. The geometry of TIP4P/2005 is maintained using the {\sc tip4} option for the choice of water model. We made sure the input TIP4P/2005 parameters are consistent with the Lorenz-Berthelot mixing rule used in simulations with NAMD. 

The lengths and relevant details of the simulations are collected in Tables~\ref{tb:namdsvr} through~\ref{tb:namdtip4pd} below. The simulations were performed in mixed precision arithmetic on hybrid CPU/GPU compute nodes in the GPU-resident mode, except for the run noted in Table~\ref{tb:namdlangevinCPU}, which was exclusively performed in double precision on CPUs. 

\setlength{\tabcolsep}{12pt}
\newcommand{\TT}[2]{\ensuremath{{#1}\pm{#2}}}
\ctable[
	mincapwidth=\textwidth,
	caption={Simulations with the CSVR thermostat/Monte Carlo barostat. The total length of the simulation is $N_t$ (in ns), of which the first 2.5~ns were set aside for equilibration. For each $\delta t$, we probed 16 temperatures from 257~K to 318~K. For cases in which not all 16 temperatures were propagated for the same length, we note a range of simulation times. The stochastic rescaling period is 1~ps. The volume is sampled every $N_v$ time-steps. Our earlier study\cite{asthagiri:cs2025} confirms that the choice for volume sampling is conservative. Configurations were saved every 500 steps for analysis.},
	label=tb:namdsvr,
	pos=h,
	captionskip=-1.ex
	]	
{c c c }
{}
{
\FL
 $\delta t$ (fs)  &  $N_t$ (ns)        &    $N_v$ (time-steps)    \ML
0.25                 & 10.28 -- 10.71     &   160             \NN[-1ex]
0.5                   &  10.57 -- 10.73    &    80              \NN[-1ex]
2.0                   &  34.70                 &    20              \NN[-1ex]
4.0                   &  33.00                 &    20              \LL
}

\ctable[
	mincapwidth=\textwidth,
	caption={Simulations with the Langevin thermostat/Langevin barostat.  For the thermostat, the Langevin damping is 1 ps$^{-1}$, and for the barostat the piston period is 200~fs with a decay of 100~fs. Rest as in Table~\ref{tb:namdsvr}.},
	label=tb:namdlangevinGPU,
	pos=h,
	captionskip=-1.ex
	]	
{c c}
{}
{
\FL
 $\delta t$ (fs)  &  $N_t$ (ns)         \ML
0.25                 &  7.69     \NN[-1ex]
2.0                   &  13.20   \NN[-1ex]
4.0                   &  26.40    \LL
}

\setlength{\tabcolsep}{12pt}
\ctable[
	mincapwidth=\textwidth,
	caption={Double precision CPU-only simulations with the Langevin thermostat/Langevin barostat.  The first 0.25~ns of simulation was set aside for equilibration.  Since the simulations were exclusively performed on CPUs, the sampling length is necessarily limited.  Rest as in Table~\ref{tb:namdlangevinGPU}.},
	label=tb:namdlangevinCPU,
	pos=h,
	captionskip=-1.ex
	]	
{c c}
{}
{
\FL
 $\delta t$ (fs)  &  $N_t$ (ns)         \ML
0.25                 &  0.38     \NN[-1ex]
2.0                   &  2.00   \NN[-1ex]
4.0                   &  3.60    \LL
}

\ctable[
	mincapwidth=\textwidth,
	caption={Simulations of TIP4P-D with the CSVR thermostat/Monte Carlo barostat. Rest as in Table~\ref{tb:namdsvr}.},
	label=tb:namdtip4pd,
	pos=h,
	captionskip=-1.ex
	]	
{c c c }
{}
{
\FL
 $\delta t$ (fs)  &  $N_t$ (ns)        &    $N_v$ (time-steps)    \ML
0.25                 & 4.46      &   160             \LL
}

\clearpage
\newpage

\subsection{AMBER simulations}

For setting up the simulation for TIP4P/2005, we found it convenient to edit a local copy of the {\sc leaprc.water.tip4pd} file to match the TIP4P/2005 parameters. Then using {\sc AmberTools}\cite{amber1} we built the appropriate {\sc prmtop} and {\sc inpcrd} files. The starting configuration is the same as the one we used in simulations using NAMD. The GPU-accelerated AMBER code\cite{amber,amber2,amber3,amber4,amber5} uses the mixed precision model, i.e.\ single precision forces and double precision accumulation. We used the CSVR thermostat and the Monte Carlo barostat with parameters matching those used in simulations with NAMD. 

The AMBER runs comprised two phases. In the first phase ($N_{t,1}$) we used the default PME Ewald tolerance ({\sc dsum\_tol}) and a PME grid of dimension $128\times 128\times 128$. In the subsequent production phase ($N_{t,2}$ ), we tightened the tolerance at the real space cutoff to 10$^{-7}$ and changed the grid dimensions to $108\times 108\times 108$ to match the choice in the runs with NAMD.  

\ctable[
	mincapwidth=\textwidth,
	caption={Simulations with the CSVR thermostat and the Monte Carlo barostat within AMBER. Rest as in Table~\ref{tb:namdsvr}.},
	label=tb:ambersvr,
	pos=h,
	captionskip=-1.ex
	]	
{c c c c}
{}
{
\FL
 $\delta t$ (fs)  &  $N_{t,1}$ (ns)         &   $N_{t,2}$ (ns)    &   $N_v$ (time-steps)    \ML
0.25                 &  5.35 -- 10.30           &       2.5               &   160             \NN[-1ex]
4.0                   &  40.0                     &      40.0               &   20               \LL
}

\clearpage
\newpage

\subsection{GROMACS simulations}

To construct the initial {\sc gro} configuration file, we used the same starting configuration as used with NAMD. We found it convenient to create a local copy of {\sc TIP4P2005.itp} by carefully editing the available {\sc TIP4PEW.itp} file within the {\sc charmm} forcefield structure. We use  the Lorenz-Bertholet mixing rule ({\sc comb-rule}=2),  to match the simulations with NAMD. 

For simulations with GROMACS\cite{gromacs2,gromacs3}, we use the Velocity Verlet integrator ({\sc md-vv}) together with the Nos\'e-Hoover thermostat and the stochastic rescaling barostat. We compiled GROMACS in the default single precision mode. The NH default chain length of 10 was used as such.  For one set of calculations we used a thermostat coupling time constant of 2~ps and a barostat coupling time constant of 5~ps. For another set of calculations we used a thermostat coupling time of 1~ps and a barostat coupling time constant of 3~ps. The latter tighter coupling leads to somewhat better agreement with our reference $\delta t = 0.25$~fs simulations with NAMD.  

\ctable[
	mincapwidth=\textwidth,
	caption={Simulations of TIP4P/2005 with GROMACS using the Nos\'e-Hoover thermostat and the stochastic rescaling barostat. Both sets of calculations for the two different thermostat/barostat coupling times were for the same length of time. Rest as in Table~\ref{tb:namdsvr}.},
	label=tb:gromacs,
	pos=h,
	captionskip=-1.ex
	]	
{c c}
{}
{
\FL
 $\delta t$ (fs)  &  $N_t$ (ns)    \ML
0.25                 &   3.56            \NN[-1ex]
2.0                  &   16.50             \NN[-1ex]
4.0                   &   15.00               \LL
}

\clearpage
\newpage

\subsection{OpenMM simulations}

The initial structure for runs with OpenMM\cite{openmm,openmm8} was the same as the one for simulations with NAMD. Within OpenMM, we used an {\sc EwaldErrorTolerance} of $10^{-6}$. Care was taken to ensure we apply long-range LJ corrections to be consistent with our simulations using NAMD. For the Nos\'e-Hoover thermostat, we used a collision frequency of 1 ps$^{-1}$. We retained the default OpenMM choice for chain length (3), the number of steps in the multiple time-step integrator (3), and the number of terms in the Yoshida-Suzuki decomposition (7). For the Monte Carlo barostat, we used a volume sampling period of 160 steps (for $\delta t = 0.25$~fs) and 20 steps (for $\delta t = 4.0$~fs). 

\begin{figure}[h!]
\includegraphics{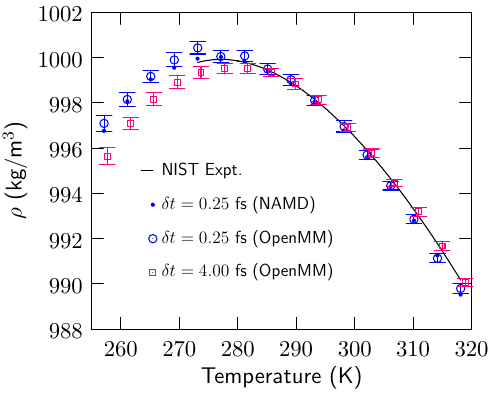} 
\caption{Predictions of $\rho$-vs-$T$ (average simulation temperature) using OpenMM. The simulated behavior at 0.25~fs is consistent with 
results from NAMD an entirely different thermostat/barostat combination. Notice also that while the density behavior might appear acceptable for $\delta t = 4$~fs, the average temperature of the simulation is clearly shifted, indicating a deviation between the setpoint and average temperatures despite the densities appearing reasonable. (Cf. Figure 4 in the main text.) The indicated standard error of the mean is at the $2\sigma$ level. The uncertainty for simulations
with NAMD is about the size of the symbol.}
\label{fg:sopenmm}
\end{figure}

Using a multiple time step algorithm factoring each step into subparts using the Yoshida-Suzuki decomposition\cite{yoshida:1990,omelyan:cpc02,leimkuhler:2015} includes force information from more 
time-step slices than what the raw step-size $\delta t$ would indicate \cite{MTTK:1996}. Such an approach perhaps better captures the frictional effects that determine the translational and rotational relaxation\cite{asthagiri:jctc2024a} than a comparable single-step Velocity Verlet method, and perhaps it is for this reason that the difference between $\delta t = 0.25$~fs and $\delta t=4.0$~fs is not as high as in the other plots shown. (But note that the distinct deviation between the average and setpoint temperatures in Fig.~\ref{fg:sopenmm}.) It must be kept in mind that the MTS approach with the Yoshida-Suzuki factorization is also considerably more computationally demanding, forcing us to reduce the system size by 1/8$^{\rm{th}}$ and also cut back on the sampling time.
Better understanding the impact on dynamical relaxation in water with higher order integrators is left for future studies. 

\ctable[
	mincapwidth=\textwidth,
	caption={Simulations of TIP4P/2005 with OpenMM using the Nos\'e-Hoover thermostat and the Monte Carlo barostat. The simulation system comprises 4096 water molecules. Rest as in Table~\ref{tb:namdsvr}.},
	label=tb:gromacs,
	pos=h,
	captionskip=-1.ex
	]	
{c c c}
{}
{
\FL
 $\delta t$ (fs)  &  $N_t$ (ns)        &    $N_v$ (time-steps)    \ML
0.25                 & 3.91 -- 3.94.         &   160             \NN[-1ex]
4.0                   &  11.9 -- 12.00       &    20              \LL
}

\newpage
\section{Supplemental results}

\subsection{NAMD CPU-vs-GPU}
\begin{figure}[h!]
\includegraphics[scale=0.9]{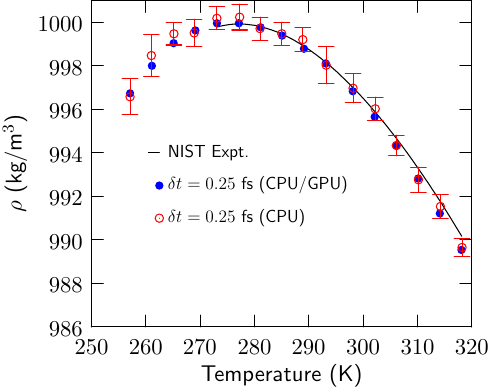}\\
\includegraphics[scale=0.9]{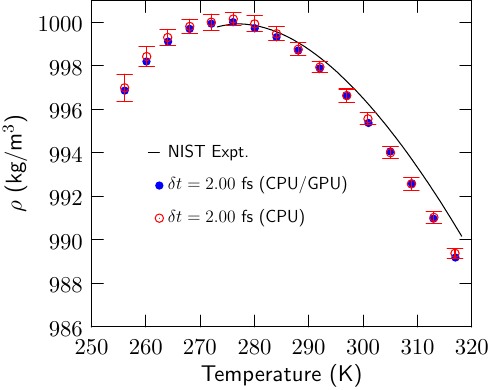}\\
\includegraphics[scale=0.9]{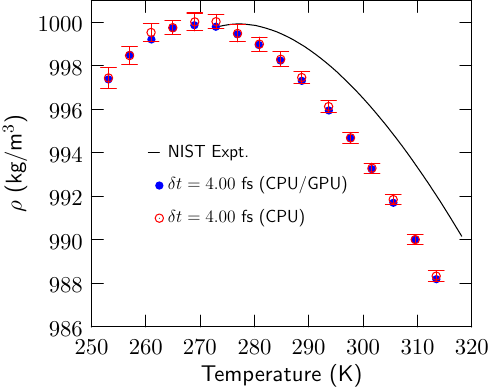}
\caption{Density-temperature behavior using the Langevin thermostat/barostat within NAMD. Within simulation uncertainties, the results using mixed precision CPU/GPU calculations agree with those using double precision CPU-only calculations.}
\end{figure}

\newpage 
\subsection{$\rho$-vs-$T$: Comparison of NAMD with AMBER}

\begin{figure}[h!]
\includegraphics{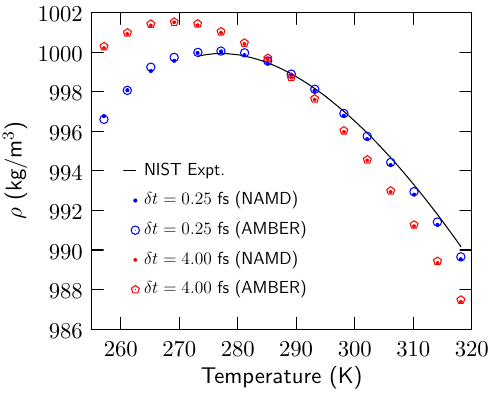}
\caption{Consistency between results obtained using AMBER and NAMD for the same thermostat/barostat combination.  The temperature denotes the average simulation temperature. Filled symbols are those from NAMD, with symbol size about 2 standard error of the mean. Open symbols are those from AMBER; the standard error is about how much we obtain with NAMD.}
\end{figure}

\newpage
\subsection{$\rho$-vs-$T$: Comparison of NAMD with GROMACS}
\begin{figure}[h!]
\includegraphics[scale=0.95]{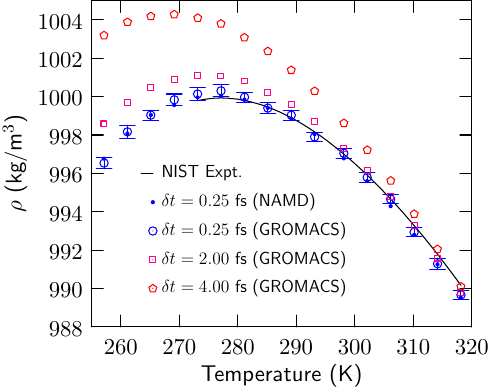} \includegraphics[scale=0.95]{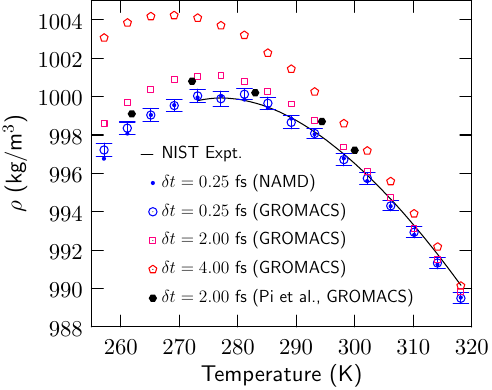}
\caption{Predictions of $\rho$-vs-$T$ (average simulation temperature) using GROMACS. The simulated behavior at 0.25~fs is consistent with results from NAMD
using an entirely different thermostat/barostat combination. Filled symbols are those from NAMD. Open symbols are those from GROMACS. The $2\sigma$ standard error of the mean is shown only for the $\delta t = 0.25$~fs case. The uncertainties for the other cases are comparable to the symbol size. \underline{Left  panel}:  Thermostat coupling time of 1~ps and a barostat coupling time constant of 3~ps \underline{Right panel}: Thermostat coupling time constant of 2~ps and a barostat coupling time constant of 5~ps. The black filled circles are results from Ref.\ \citenum{pi:tip409}, where the authors use the Nos\'e-Hoover thermostat with a 2~ps relaxation and a Parrinello-Rahman barostat with a 2~ps relaxation; note the consistency with results obtained in this work using a 2~fs time-step. The identification of a TMD of $277\pm 3$~K in Ref.\ \citenum{pi:tip409} is likely a fortuitous  outcome of using a polynomial interpolant to locate the TMD.}
\end{figure}

\newpage
\subsection{$\rho$-vs-$T$: Comparison of NAMD with OpenMM}

\clearpage
\newpage
\bibliography{refs.bib}